\documentclass[twocolumn]{aastex631}

\usepackage[utf8]{inputenc}
\usepackage[T1]{fontenc}
\usepackage{mathrsfs} 
\usepackage{amsmath,amsthm,amssymb,amsfonts}
\usepackage{listings,graphicx}
\usepackage{hyperref}
\usepackage{natbib}
\usepackage{enumitem}                           
\usepackage{mathtools}                          
\usepackage{environ}                            
\usepackage{color}

\newcommand{\rev}[1]{{\color{black}#1}}

\defcitealias{lovelace_rossby_1999}{L99}
\defcitealias{ono_parametric_2016}{O16}

\newcommand{\comment}[1]{} 

\setenumerate[0]{label=(\alph*)}

\NewEnviron{gatheralign}{
\begin{gather}
    \begin{aligned}
        \BODY
    \end{aligned}
\end{gather}
}

\let\oldgatheralign\gatheralign
\let\oldendgatheralign\endgatheralign
\renewenvironment{gatheralign}
  {\linenomathNonumbers\oldgatheralign}
  {\oldendgatheralign\endlinenomath}

\newcommand{\dcasesb}{\begin{dcases}}
\newcommand{\dcasese}{\end{dcases}}

\newcommand{\lp}{\left(}
\newcommand{\rp}{\right)}
\newcommand{\lb}{\left[}
\newcommand{\rb}{\right]}

\newcommand{\lbr}{\left\{}
\newcommand{\rbr}{\right\}}

\submitjournal{ApJL: 02/01/23 (Accepted: 03/03/23)}
\shorttitle{Constraining Dust Structures with the RWI}
\graphicspath{{./}{figures/}}

\begin{document}

\title{On the Origin of Dust Structures in Protoplanetary Disks:\\
Constraints from the Rossby Wave Instability}
\author[0000-0003-4703-2053]{Eonho Chang}
\affiliation{Graduate Interdisciplinary Program in Applied Mathematics, University of Arizona, Tucson, AZ 85721, USA}
\affiliation{Department of Astronomy and Steward Observatory,  University of Arizona, Tucson, AZ 85721, USA}

\author[0000-0002-3644-8726]{Andrew N. Youdin}
\affiliation{Department of Astronomy and Steward Observatory,  University of Arizona, Tucson, AZ 85721, USA}
\affiliation{The Lunar and Planetary Laboratory, University of Arizona, Tucson, AZ 85721, USA}

\author[0000-0001-7671-9992]{Leonardo Krapp}
\affiliation{Department of Astronomy and Steward Observatory,  University of Arizona, Tucson, AZ 85721, USA}



\begin{abstract}
High resolution sub-mm observations of protoplanetary disks with ALMA have revealed that  dust rings are common in large, bright disks. The leading explanation for these structures is dust-trapping in a local gas pressure maximum, caused by an embedded planet or other dynamical process. Independent of origin, such dust traps should be stable for many orbits to collect significant dust.  However, ring-like perturbations in gas disks are also known to trigger the Rossby Wave Instability (RWI).  We investigate whether axisymmetric pressure bumps can simultaneously trap dust and remain stable to the RWI.  The answer depends on the thermodynamic properties of pressure bumps.  For isothermal bumps, dust traps are RWI-stable for widths from ${\sim}1$ to several gas scale-heights.  Adiabatic dust traps are stable over a smaller range of widths. For temperature bumps with no surface density component, however, all dust traps tend to be unstable. Smaller values of disk aspect ratio allow stable dust trapping at lower bump amplitudes and over a larger range of widths.  We also report a new approximate criterion for RWI. Instability occurs  when the radial oscillation frequency is $\lesssim75$\% of the Keplerian frequency, which differs from the well-known Lovelace necessary (but not sufficient) criterion for instability. Our results can guide ALMA observations of molecular gas by constraining the resolution and sensitivity needed to identify the pressure bumps thought to be responsible for dust rings.
\end{abstract}



\section{Introduction}
High resolution observations of protoplanetary disks by the Atacama Large Millimeter/submillimeter Array (ALMA) have revealed a variety of substructures, including axisymmetric features such as rings and gaps, as well as non-axisymmetric vortex-shaped or crescent-shaped traps  \citep{van_der_marel_major_2013, HLtau2015,andrews_disk_2018}. 

These regions of enhanced continuum emission correspond to locations where dust has concentrated and/or become heated.  The leading hypothesis is that these structures form when dust drifts into local maxima in gas pressure \citep{Whipple1972, pinilla_particle_2017}.  However, alternate explanations have been proposed, including: the concentration of dust by a ``secular" gravitational instability of the dust layer \citep{youdin_2011_sgi, Takahashi2016}; a thermal shadowing instability of the disk \citep{Ueda2021}; and changes in dust properties near condensation fronts, i.e. ``snow lines" \citep{zhang_evidence_2015,Okuzumi2016}.  These mechanisms, and related ones, are reviewed in \cite{Bae2022PP7}.

For the leading hypothesis of dust concentration in pressure bumps, the pressure bumps could have a planetary or non-planetary origin.  The outer edge of planet-carved gaps can trap dust in a pressure maxima \citep{Paardekooper2004, lyra_standing_2009, Pinilla2012}.
Without planets, a variety of dynamical mechanisms could also create a pressure bump.  These include:  zonal flows arising in magneto-rotational turbulence \citep{Johansen2009, Krapp2018}; dead zone boundaries \citep{lyra_embryos_2008, Ruge2016}; magnetized disk winds \citep[][]{Suriano2017,Riols2019}; and the vertical shear instability \citep{nelson_linear_2013, lin_cooling_2015, Flock2017}.

Therefore, identifying whether or not dust is concentrated in gas pressure maxima will aid our understanding of the nature of dust substructures and their role in planet formation.  ALMA observations of molecular lines, especially of CO, combined with chemical models, constrain the mass and temperature distribution of disk gas \citep{oberg_molecules_2021}.  However the spatial and velocity resolution is not  sufficiently high in current observations to clearly confirm or rule out gas pressure maxima as the source of dust structures.  The goal of this letter is to theoretically constrain the properties of gas structures that can trap dust, to aid the planning and interpretation of ALMA observations.  Specifically, we require that dust-trapping pressure bumps be dynamically stable.

\rev{Specifically} the Rossby Wave Instability (RWI, \citealp[][ hereafter \citetalias{lovelace_rossby_1999}]{lovelace_rossby_1999})  is triggered by narrow, ring-like gas structures.  There are two main non-linear outcomes to the RWI.  First, the RWI can trigger the formation of vortices \citep{Li2001}, whether the initial ring-like perturbation was formed by a planet \citep{Koller2003} or by another source such as a dead zone boundary \citep{Varniere2006}.   Second, \rev{after vortices decay, ring-like structures spread out to a RWI-stable state}  
\citep{hammer_kratter_2017}.  In either case, an axisymmetric pressure bump should not persist in a RWI-unstable state.  

Thus if the dust rings observed by ALMA are caused by pressure trapping, the pressure bump should be RWI-stable, or at most marginally unstable.  In this letter, we use this constraint to place limits on the amplitudes and widths of gas bumps that could produce observed dust rings. \rev{We believe that this work provides the first systematic comparison of the conditions for pressure trapping and hydrodynamic stability.  Most similarly, \citet{yang_rayleigh_2010} considered the effect of the axisymmetric Rayleigh instability on gas bumps and steps.  However the non-axisymmetric RWI is more readily triggered (see Section \ref{subsec:RWIstabapprox}) and thus places more stringent constraints on gas rings.  Moreover, neither that work, nor other previous works (to our knowledge), have addressed the main question we are asking: \emph{Which stable gas rings can also trap dust?}}

In Section \ref{sec:methods}, we describe our model of a disk with a bump and the methods of our stability analysis. Section \ref{sec:results} presents our results for the properties of stable, dust-trapping rings. In Section \ref{sec:discussion}, we discuss the implications and possible extensions of our work.

\section{Methods}
\label{sec:methods}

\subsection{Disk-bump model}
\label{subsec:model}
We consider a set of simple, but flexible models of a protoplanetary disk with a bump.   
These axisymmetric models have surface mass density $\Sigma$ and (vertically isothermal) temperature $T$ that vary with radius $R$ as
\begin{gatheralign}\label{eq:diskmodel}
    \Sigma(R) &\equiv 
    \Sigma_0 \lb \lp\frac{R}{R_0}\rp^n +  A_\Sigma g(R-R_0, W) \rb ,\\
    T(R) &\equiv 
    T_0\lb \lp\frac{R}{R_0}\rp^q + A_T g(R-R_0, W) \rb ,
\end{gatheralign}
where the background disk slope is given by the exponents $n$ and $q$. The bumps are Gaussian-shaped, with $g(R-R_0, W) \equiv \exp [-(R-R_0)^2/(2W^2)]$, centered on $R_0$ with width $W$.  The bump amplitudes are $A_\Sigma$ and $A_T$.\footnote{Unlike some previous works \citep[e.g.][]{lovelace_rossby_1999,li_rossby_2000}, our bump is not multiplied by the background power-law.  This choice allows our large amplitude bumps to be independent of the background slope.} In the absence of a bump, the disk would have the background values, $\Sigma_0$ and $T_0$, at $R_0$.

We assume an ideal gas, with pressure $P \propto \rho T$ for mass density $\rho$.  The structure of $P$ and $\rho$ with vertical distance $z$ from the disk midplane follows from hydrostatic balance.  We neglect disk self-gravity and use the vertical gravitational acceleration of a thin disk, $g_z = - \Omega_\mathrm{K}^2 z$ with the Keplerian frequency $\Omega_\mathrm{K} \propto R^{-3/2}$.  The midplane density and pressure can then be written
\begin{gatheralign}
\rho_{\rm m} &= \frac{\Sigma}{\sqrt{2 \pi} H}\, , \\
P_{\rm m} &= \frac{\Sigma}{\sqrt{2 \pi}}H \Omega_\mathrm{K}^2\, .
\end{gatheralign}
For the gas scale-height $H$, we specify the aspect ratio
\begin{gatheralign}
    h \equiv \frac{H}{R} = h_0 \sqrt {\frac{T}{T_0} \frac{R}{R_0}}\, .
\end{gatheralign}
The midplane pressure $P_\mathrm{m}$ is used to determine the location of dust traps, because dust will accumulate at a pressure maximum with $d P_\mathrm{m}/dR = 0$ \citep{Whipple1972, houches10}.
For the RWI analysis, we use a height integrated disk model, where the relevant pressure is $P_\mathrm{H} = \int_{-\infty}^\infty P dz = P_\mathrm{m} \sqrt{2\pi} H$. 
We henceforth drop the subscript ``${\rm m}$'' from midplane values for convenience.

Choosing scaled units of $R_0, \Sigma_0$ and $\Omega_0 \equiv \Omega_\mathrm{K}(R_0)$, we specify our model by six dimensionless parameters $n, q, h_0, A_\Sigma, A_T$ and $W/H_0 = W/(R_0h_0)$.  We further fix  an effective, height-integrated adiabatic index, $\Gamma = 4/3$,  which is approximately equivalent to a standard (3D) adiabatic index of $7/5$, appropriate for diatomic molecules  \citep{ostriker92}.

\begin{figure}
    \centering
    \includegraphics[width=0.45\textwidth]{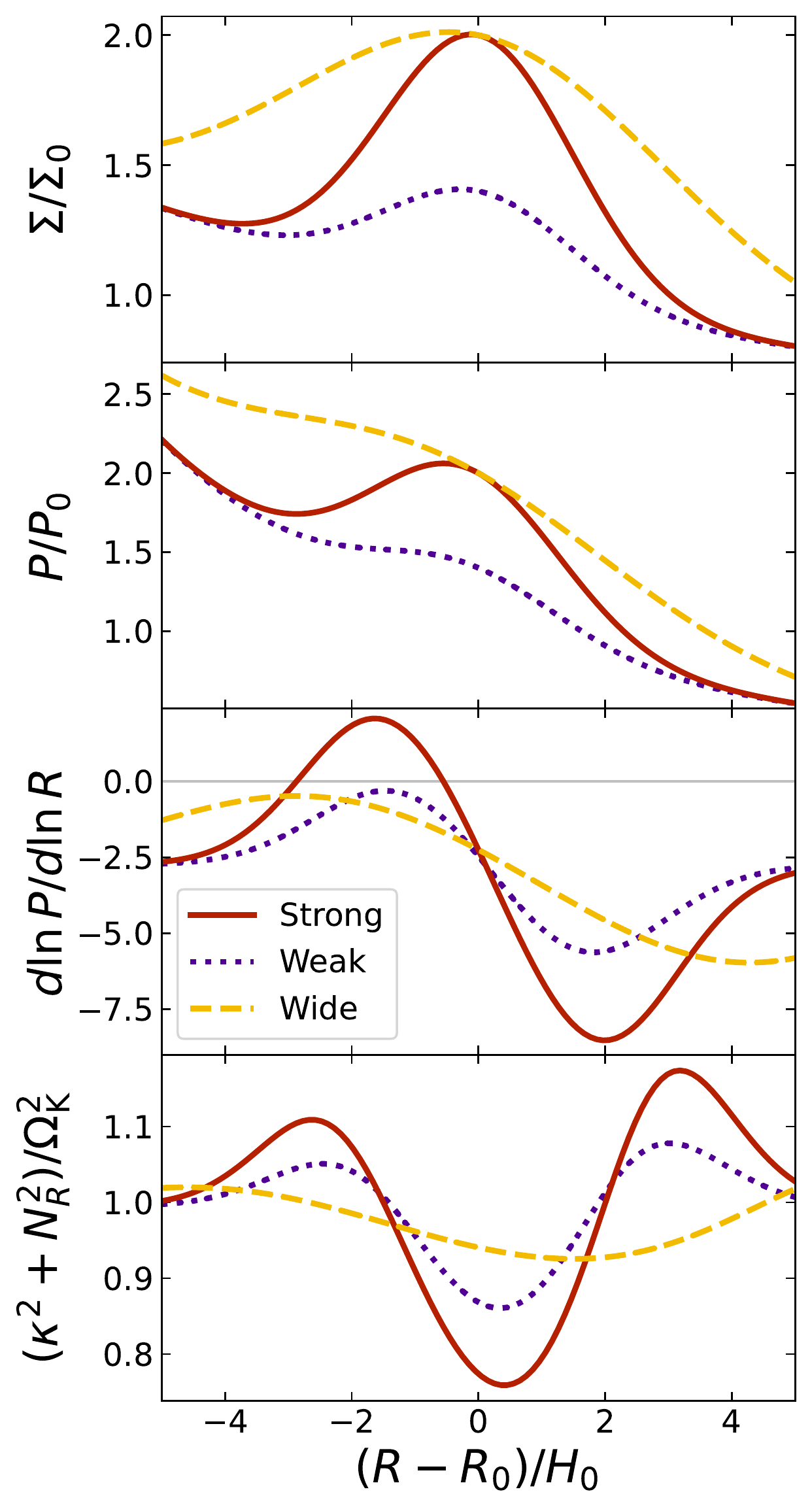}
    \caption{The behavior of three different bumps in our fiducial disk model:  ``Strong'' (which is also narrow), ``Weak'' and ``Wide'' with $A_\Sigma=(1, 0.4, 1),\, W/H_0 =(1.5, 1.5, 3)$, respectively. 
    \emph{Top two panels}: Surface density, $\Sigma$, and midplane pressure, $P$, in the vicinity of the bump. \emph{Third panel}: The pressure gradient, which only becomes positive in the ``Strong" case, indicating a dust-trapping pressure maximum.  \emph{Bottom panel}: the disk's squared radial oscillation frequency, relative to the squared Keplerian frequency.  These radial frequencies are related to disk stability.}
    \label{fig:r_profiles}
\end{figure}

Our fiducial model includes a bump in surface density but not in temperature (i.e.\ $A_\Sigma>0,\,A_T = 0$).  The parameters of the fiducial case are $n = -1$, $q = -0.5$, $h_0 = 0.05$ and $A_T = 0$, with different choices of $A_\Sigma$ and $W/H_0$. In Section \ref{subsec:background} we explore deviations from these background disk parameters, and in Section \ref{subsec:heatbump} we consider heated bumps with $A_T > 0$.

Our choices of background disk parameters span most theoretical expectations, as well as observational constraints, especially in the $R_0 \simeq 50{-}100$ AU region where ALMA has observed prominent dust rings, e.g.\ with the DSHARP survey \citep[][]{dullemondDSHARP2018}. Gas parameters are constrained by ALMA observations of molecular lines plus thermo-chemical modelling.   \citet{zhang_molecules_2021} fit ALMA MAPS survey data to a disk model with an exponentially truncated power-law disk in $\Sigma$.  The local slope $n = d \ln \Sigma/ d \ln R$ of those models vary from $-1$ to $-2$ at $R_0 \simeq 50{-}100$ AU. Their fits also give values from $q\simeq-0.3$ to $-0.8$ throughout the disk, and values $h_0\simeq 0.03$ in the inner disk ($R_0\simeq 10$ AU) and $h_0\simeq 0.1$ in the outer disk ($R_0\simeq 150$ AU),  consistent with our choices.  

Figure\ \ref{fig:r_profiles} illustrates examples of ring-like bumps in our fiducial disk model.  The red solid curves represent a relatively strong and narrow bump with  $A_\Sigma = 1, W/H_0 = 1.5$. A weaker bump (purple dotted curves) and a wider bump (yellow dashed curves) are shown for comparison. The top two panels show the bumps in surface density and in midplane pressure, respectively.  The bumps appear less prominent in pressure than in $\Sigma$ due to the steeper background, power-law slope of pressure.  The third panel shows the logarithmic pressure gradient,  i.e.\ the local power-law slope.  Both the weak and wide bumps have negative slope everywhere, implying that dust drift is directed solely inward due to sub-Keplerian gas orbits. For the strong bump, the slope briefly becomes positive and dust can collect in the local pressure maximum. 

In the weak and wide cases, the reduction of inward drift speeds --- where  radial pressure gradients are weak, but still negative --- would increase the dust density as a traffic jam effect \citep{carrera_protoplanetary_2021}.  We focus on the stronger dust concentrations that occur in local pressure maxima. 

The bottom panel of Figure\ \ref{fig:r_profiles} shows the disk's radial oscillation frequency squared, a combination of the epicyclic frequency, $\kappa$, and radial buoyancy frequency, $N_R$, computed as in \citetalias{lovelace_rossby_1999}.
 Negative values of $\kappa^2 + N_R^2$ would imply instability by the Solberg-H\o iland criterion \citep[]{lin_cooling_2015}.  
 While this quantity remains positive, it is reduced near $R_0$ by the pressure bump.  As we show in Section \ref{subsec:RWIstabapprox}, even a partial reduction could trigger the RWI. Specifically, we find $\kappa^2 + N_R^2 \lesssim 0.6 \Omega_\mathrm{K}^2$ somewhere to be an approximate criterion for the RWI.   The strong and narrow bump which traps particles also causes the largest reduction of  $\kappa^2 + N_R^2$, which makes it closer to triggering the RWI.  This particular bump turns out to be stable to the RWI, and the goal of this letter is to explore systematically when this is true in different circumstances.

\subsection{RWI Stability Analysis}\label{subsec:RWImeth}



\rev{This work studies the RWI stability of disks with bumps, as parameterized in Equation \ref{eq:diskmodel}, for the purpose of comparing to the conditions for trapping dust.  However, 
determining RWI stability} requires a numerical calculation as no general, analytic criterion for the RWI exists. 
The well-known Lovelace criterion \citepalias{lovelace_rossby_1999} provides a necessary, but not sufficient, condition for instability. 
 
\rev{Many works, starting with \citet{li_rossby_2000}, have computed the linear growth of the RWI.  Of these,} \citet[][hereafter \citetalias{ono_parametric_2016}]{ono_parametric_2016} performed the most thorough analysis to date of the RWI stability boundary, finding the amplitudes and widths needed for bumps to trigger instability.   \citetalias{ono_parametric_2016} also considered gaps and steps, which we ignore here, partly because the 
\rev{results are similar, and also because we wish to more thoroughly examine the bump case in this initial study.}




\rev{Comparing to \citetalias{ono_parametric_2016}, our work has two key distinctions.  First and foremost, we are comparing to the conditions for dust trapping.  Second, \citetalias{ono_parametric_2016} considered disks with a flat background ($n = q = 0$) and a barotropic equation of state ($T \propto \Sigma^{\Gamma-1}$).  We study non-barotropic disks as well, to consider a wider, and more realistic, range of background disk slopes and also to compare bumps in surface density to bumps in temperature.  We thus note that --- as a means to the end of better understanding dust trapping pressure bumps --- our results build on \citetalias{ono_parametric_2016} by performing the most thorough investigation to date of the RWI stability boundary for non-barotropic disks.}



Our stability analysis uses the original height-integrated, linearized equations of \citetalias{lovelace_rossby_1999}.  \rev{These equations are non-barotropic} which allows for disk entropy gradients, and thus radial buoyancy.  \rev{Furthermore, these equations} assume adiabatic perturbations, i.e.\ no cooling.  
Specifically, we solve the ODE of Equation (10) in \citetalias{lovelace_rossby_1999}, which describes the behavior of linear perturbations to an equilibrium disk model.  We use our disk model, Equation \ref{eq:diskmodel}, as the equilibrium, using the $P_\mathrm{H}$ as the relevant, height-integrated pressure.

\rev{We solve the governing ODE using the same method and boundary conditions as those of \citetalias{ono_parametric_2016}, described in their Appendix.  The wave frequencies, $\omega$, and RWI growth rates, $\gamma$, are found as the complex eigenvalues of the resulting linear system, using M\"uller's method. For all linear stability calculations, we use $N=3000$ grid points uniformly spaced in the radial domain $R\in[0.3R_0,3R_0].$
As a check on our calculations, we reproduced the stability boundary that \citetalias{ono_parametric_2016}  found for their bump cases, (iii) and (iv).

While the RWI has been been analyzed in 3D \citep{Meheut2010, lin_non-barotropic_2013}, even the linear calculations are considerably computationally intensive.  Fortunately, growth rates appear similar in 2D and 3D \citep{Meheut2012}, though an investigation of the RWI stability boundary in 3D is left to future work.}

A technical difficulty in studying the marginal stability to the RWI is that when $\gamma = 0$, the governing ODE is singular at corotation.\footnote{Corotation is where the wave's pattern speed, $\omega/m$, matches the disk's orbital frequency, $\Omega(R)$, and as a result the Doppler-shifted frequency vanishes: $\Delta \omega \equiv \omega - m \Omega(R) = 0$ \citep{tsang08}. In practice, the corotation radius is located near the center of the bump.}
To avoid the singularity, we are restricted to finding solutions near marginal stability, and our main results are for $\gamma = 5 \times 10^{-3}\Omega_0$.  Such growth rates are sufficiently slow for two reasons.  First, since RWI growth rates increase rapidly away from the stability boundary and the precise threshold chosen for $\gamma/\Omega_0 \ll 1$ has little effect on the inferred boundary.
Second, linear growth that is slower than hundreds of orbits is unlikely to be astrophysically relevant, as it becomes a significant fraction of the disk lifetime in the outer disk and unlikely to be the dominant dynamical effect.

\citetalias{ono_parametric_2016} were able to remove the corotation singularity of order $\mathcal O(1/\Delta\omega)$ for marginally stable states (see their Section 5.2) and confirm that the $\gamma = 0$ and small $\gamma$ boundaries were indistinguishable.  Their  technique  works for barotropic disks with no radial entropy gradients, but could not be applied to our non-barotropic model.  Moreover, for our non-barotropic case, the corotation singularity is of higher order, $\mathcal O(1/\Delta \omega^{2})$, which means that we require higher grid resolution to solve for a given small growth rate.

\begin{figure*}
    \centering
    \includegraphics[width=\textwidth]{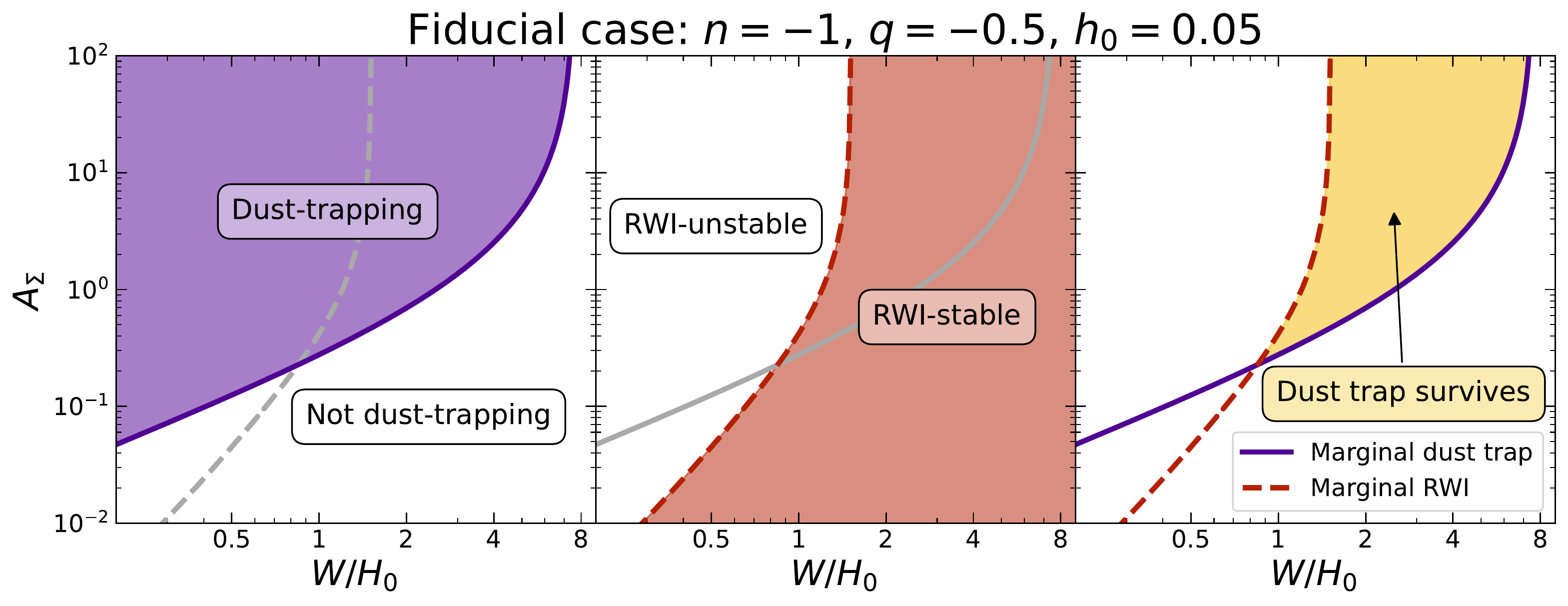}
    \caption{The ability of ring-like surface density bumps --- with amplitude, $A_\Sigma$, and width relative to local gas scale-height, $W/H_0$ --- to trap dust and/or trigger hydrodynamic instability is shown, for our fiducial disk model. 
    \emph{Left:} the shaded region, above and to the left of the purple solid curve, denotes bumps that reverse the sign of the background pressure gradient and thus can trap dust. 
    \emph{Middle:} the shaded region, below and to the right of the red dashed curve, denotes bumps that are stable to the RWI, or with a low growth rate.
    \emph{Right:} the yellow shaded region shows where the previous shaded regions overlap, giving a region where the bumps can both trap dust and not trigger significant RWI.  Gas rings with these properties could produce observed dust rings.  Bumps with amplitude and width outside the shaded region will fail to trap dust and/or be modified by the RWI.  } 
    \label{fig:ref_case}
\end{figure*}

In finding the stability boundary, we fix the azimuthal wavenumber to $m = 1$. This mode was found to be the most unstable near marginal stability by \citetalias[][]{ono_parametric_2016}, in the sense of giving instability for the smallest bump amplitudes, at a given width.  We also investigated whether this result held for our models, which are non-barotropic and include smaller values of $h_0$. With our fiducial model, we confirmed that $m=1$ is the fastest growing mode as $\gamma \rightarrow 0$. 
However near the $\gamma=5\times10^{-3}\Omega_0$ threshold used in this work, $m>1$ modes can be the fastest growing, but only for narrow widths ($W/H_0\lesssim0.5$).
The RWI stability boundary would thus move to somewhat smaller amplitudes at narrow widths if $m>1$ modes were included.  But this shift would not affect our results because amplitudes are already too low for dust-trapping in this region (as shown in Figure \ref{fig:ref_case}).  

\section{Results}
\label{sec:results}

We find the properties of ring-shaped bumps in gas disks which could explain the bright dust rings observed by ALMA, because they can both trap dust in a pressure maximum, and remain stable to the RWI.  A dust trap should be stable for hundreds of orbital times ($\Omega_{\rm K}^{-1}$) for significant amounts of dust to accumulate.

The radial drift timescale is at least ${\sim}1/h_{\rev{0}}^2\Omega_{\rev{0}}^{-1} \simeq 400 \Omega_{\rev{0}}^{-1}$ if dust over a large radial scale ${\sim}R_0$ accumulates in a ring.  Drift times are longer if particles are not of the optimum size (${\sim}$cm) at which the drag and orbital timescales match \citep{adachi1976, chiang2010}.  Despite uncertainties in grain size and ring feeding zone, this drift timescale is similar to, or longer than, the adopted ``marginal" growth timescale, $1/\gamma = 200 \Omega_{\rev{0}}^{-1}$ in our RWI analysis.  Our stability constraints would become somewhat tighter if pressure traps need to survive for even longer to accumulate dust.  However, as noted in Section \ref{subsec:RWImeth}, our results are not very sensitive to the choice of growth rate, as long as $\gamma/\Omega_{\rev{0}} \ll 1$.

Results for our fiducial disk model are in Section \ref{subsec:ref}.  We vary the background disk power-laws and aspect ratio in Section \ref{subsec:background} and then consider the effect of the temperature on the pressure bump in Section \ref{subsec:heatbump}.  Section \ref{subsec:RWIstabapprox} presents an approximate, empirical, and apparently rather general stability criterion for the RWI.

\subsection{Fiducial case}\label{subsec:ref}

Our fiducial model considers a pressure bump parameterized by choices of the bump amplitude in surface density, $A_\Sigma$, and the width, $W$.  This bump has the temperature of the background disk (i.e. $A_T = 0$) with background disk parameters given in Section \ref{subsec:model}.  

The ability of these bumps to produce pressure maxima, and thus trap dust, is shown in the left panel of Figure \ref{fig:ref_case}.  The colored region  shows that bumps with larger amplitudes and narrower widths produce dust traps.  The critical curve (in purple) shows the minimum amplitude needed for a dust trap at a given width, so that $dP/dR \geq 0$ somewhere.  At low amplitudes, the critical amplitude increases linearly with width, simply because a bump's maximum pressure gradient scales as $A_\Sigma/W$.  For $A_\Sigma \gtrsim 1$, however, the critical amplitude increases more sharply.  
 To explain this steepening, we look at the effect of the bump on $d\ln P/d\ln R$. The logarithmic gradient is the most relevant one since the pressure gradient is affected by other power-laws, such as the Keplerian rotation $\Omega_\mathrm{K} \propto R^{-3/2}$. We find the dependence of the power-law slope $d\ln\Sigma/d\ln R$ (and consequently $d\ln P/d\ln R$) on $A_\Sigma$ to diminish for $A_\Sigma\gtrsim 1$.  
 Thus, dust-trapping pressure bumps are unlikely to be wider than a few scale-heights since the required bump amplitudes would be extremely large.
 
The middle panel of Figure\ \ref{fig:ref_case} shows the amplitudes and widths  that are either unstable or stable to the RWI. Qualitatively, the RWI-unstable region consists of larger amplitudes and narrower widths, similar to the dust-trapping region.  This general similarity is the reason we investigate these regions more quantitatively.  The curve of marginal RWI-stability (red dashed) was determined numerically, as described in Section \ref{subsec:RWImeth}, and we explain its detailed shape in Section \ref{subsec:RWIstabapprox}.

To roughly explain the behavior of the RWI stability curve, we note that the instability is mainly driven by shear from the pressure gradient --- or more specifically, the gradient of the pressure gradient.  Orbital shear thus scales as $A_\Sigma/W^2$ at low amplitudes, explaining the $A_\Sigma \propto W^2$ slope at low amplitudes.  At large amplitudes, the strength of pressure gradients saturates with increasing amplitude, since $\Sigma^{-1}dP_{\rm H}/dR \propto d\ln \Sigma/dR$ with no temperature component to the bump.  Again, the weak dependence of $d\ln \Sigma/dR$ on $A_\Sigma$ at $A_\Sigma\gtrsim1$ explains the steepening of the RWI stability curve in $A_\Sigma$ vs.\ $W$ at large amplitudes.

Since the stability boundary crosses the critical curve for pressure trapping, an overlap for RWI-stable pressure bumps exists as shown in the right panel of Figure\ \ref{fig:ref_case} (shaded yellow).  If pressure bumps are the cause of ring-shaped dust structures in ALMA disks, then the bumps should lie in this wedge-shaped region.  Specifically, for the fiducial case, the region of stable dust traps occurs above a minimum bump amplitude and for a range of widths, starting at around a gas scale-height, $W \sim H_0$.  For larger amplitudes (above the minimum), the minimum width and the range of widths increases, reaching a few $H_0$.  

However, the properties of stable dust traps, and even their existence, depend on disk properties that we vary in the next subsections.

\subsection{Background disk effects}
\label{subsec:background}

\begin{figure*}
    \centering
    \includegraphics[width=\textwidth]{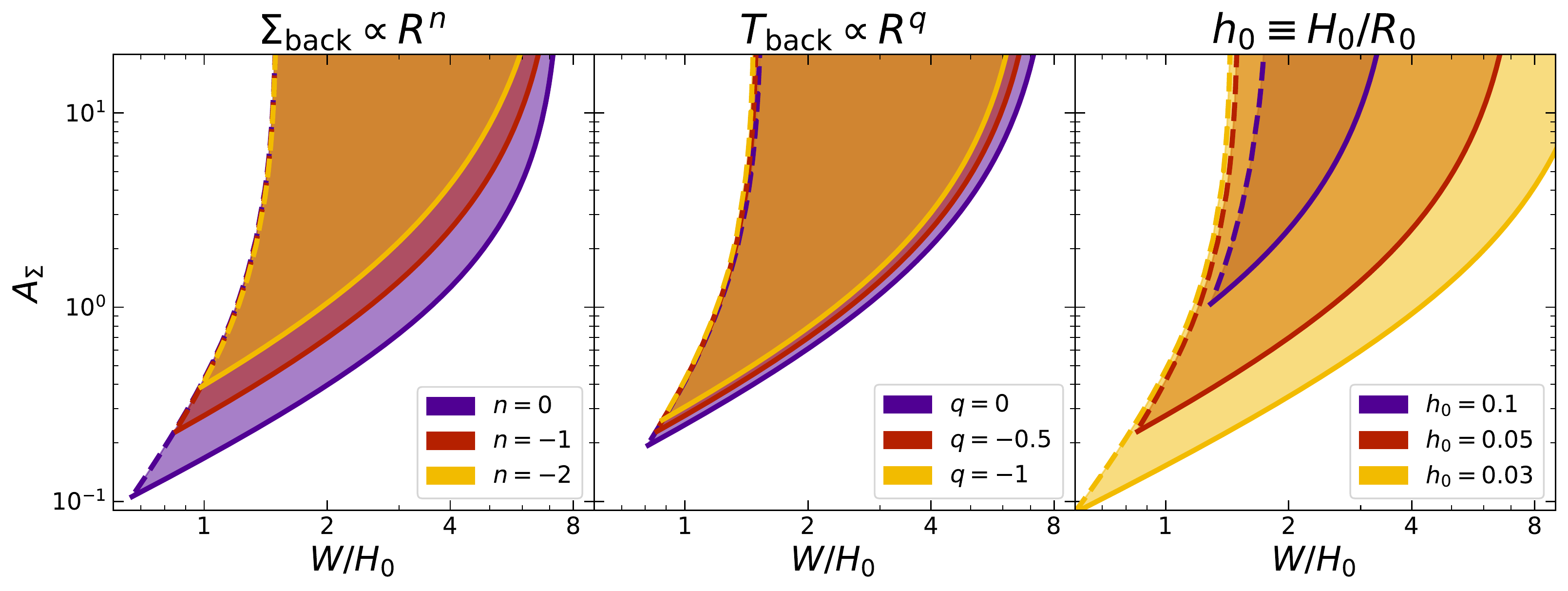}
    \caption{The effect of the background disk on the properties of stable dust-trapping rings (indicated by the shaded regions) is shown, generalizing the fiducial case shown in Figure \ref{fig:ref_case}.  The effects of slopes in the background surface density ($n$, \emph{left}) and temperature ($q$, \emph{middle}) and of the aspect ratio ($h_0$, \emph{right}) are shown.  The most significant effect is that colder disks with smaller $h_0$ can host stable dust trapping rings with smaller amplitudes, and over a greater range of widths.}
    \label{fig:eff_params}
\end{figure*}

We now probe how the region of stable dust traps depends on the properties of the background disk, namely the aspect ratio, $h_0$, and the power-laws, $n$ and $q$, for the surface density and temperature, respectively.  As with the fiducial case, we fix $A_T = 0$.

Specifically, we consider a range of values $n\in\lbr-2,-1,0\rbr,\,q\in\lbr-1,-0.5,0\rbr,\,h_0\in\lbr0.03,0.05,0.1\rbr$, which are consistent with observational expectations as discussed in Section \ref{subsec:model}.
\rev{While the flat $n=0$ and $q=0$ cases seem less realistic, they are included as an idealized control.}

The effect of the background disk slopes, $n$ and $q$, on the properties of stable dust traps is shown in the left and middle panels, respectively, of Figure\ \ref{fig:eff_params}.  Neither slope has a significant effect on the RWI stability boundary (dashed curves on the left).  This independence is expected since smooth disk gradients do not introduce significant shear or vortensity.  Both slopes do affect the pressure trapping boundary, on the right of the stable dust trapping region.  This effect is also expected since a pressure maximum involves a competition between the pressure gradients caused by the background and bump.  Thus with flatter background (e.g. $n=0$, $q=0$), smaller and wider bumps can create pressure traps.  The effect for the temperature slope $q$ appears weaker for two reasons.  First, the pressure has a weaker dependence on temperature, $P \propto \Sigma \sqrt{T}$, when the scale-height is accounted for. Second, a smaller range of $q$ is considered. 

The disk aspect ratio, $h_0$, has a strong effect, as shown in the right panel of Figure\ \ref{fig:eff_params}.  Colder, thinner disks with smaller $h_0$ have an expanded region of parameter space for stable, dust-trapping rings.  To understand the effect of varying $h_0 \equiv H_0/R_0$, it important to note that widths are plotted relative to $H_0$.  From this perspective it appears that with changing $h_0$ the RWI stability boundary changes little, while the pressure trapping boundary expands for colder disks.  However the effect is perhaps easier to understand when considering bump widths relative to the radial length-scale of the disk, $W/R_0$.  From this perspective the pressure trapping boundary does not change, since the pressure gradients of the background and bump are described by the length-scales $R_0$ and $W$, independent of $H_0$.  Meanwhile, the RWI stability boundary moves to smaller $W/R_0$ for smaller $h_0$.  This shift occurs because the strength of pressure gradients relative to Keplerian gravity scales as $h_0^2$.  Thus for a smaller value of $h_0$, bumps have to be narrower to produce the same velocity deviation. 

The fact that colder disks with smaller $h_0$ can have lower-amplitude stable pressure traps is significant. Regardless of their origin, lower-amplitude bumps should be more readily produced in these disks.

\begin{figure*}
    \centering
    \includegraphics[width=\textwidth]{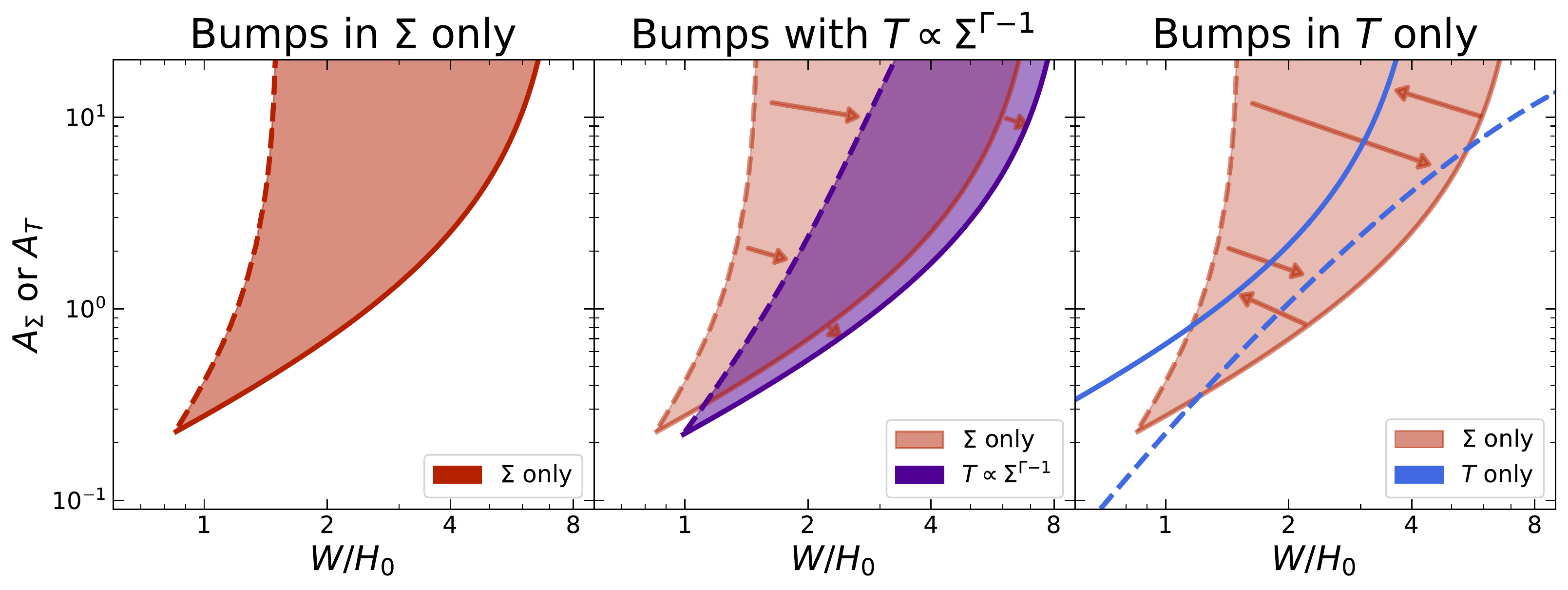}
    \caption{The effect of bump heating on the existence of stable dust traps, where the bump amplitude on the $y$-axis is $A_T$ in the right panel and $A_\Sigma$ in the left and middle panels.
    \emph{Left}: the red shaded region denotes stable dust traps in the fiducial case of isothermal bumps.  \rev{This red region is repeated more transparently in the other panels for comparison, with arrows roughly showing the directions that the boundaries change.} \emph{Middle}: the red shaded region shows stable dust traps for bumps in both surface density bump and temperature, related adiabatically. The parameter space for stable dust traps is reduced. \emph{Right}: the case of a \rev{temperature} bump with no surface density variation. There is no shaded region, as stable dust traps do not exist in this case due to the relative locations of the pressure trapping boundary (\emph{solid}) and the RWI boundary (\emph{dashed}).}
    \label{fig:eff_bumptypes}
\end{figure*}

\subsection{Effects of heated bumps}\label{subsec:heatbump}


Thus far, we have considered isothermal bumps in surface density, i.e.\ $ A_T = 0$. 
\rev{Since the thermodynamic properties of pressure bumps are not well-constrained, we consider bumps with a temperature component as well. Specifically, we examine two types of bumps: in surface density which are also heated, and in temperature alone, which \citet[][]{kim_detection_2020} showed were a possibility for the CR Cha disk.} 
For reference, the left panel of  Figure \ref{fig:eff_bumptypes} shows the properties of stable dust traps in our fiducial, isothermal model.  The other panels \rev{show the effects} of heated bumps.  The parameter space for stable dust traps is reduced or eliminated in heated bumps, as explained below.

The middle panel of Figure \ref{fig:eff_bumptypes} considers bumps with  $T\propto \Sigma^{\Gamma-1}$, as by adiabatic compression with no cooling.  For this case (only), the disk temperature (background and bump) is given not by Equation \ref{eq:diskmodel} but by the adiabatic relation $T/T_0 =  (\Sigma/\Sigma_0)^{\Gamma-1}$, with the fiducial values of $n$ and $h_0$.\footnote{The adiabatic case thus has $q = n(\Gamma -1) = -1/3$ far away from the bump.  We showed in Figure\ \ref{fig:eff_params} that modest changes to $q$ have little effect on our results.}

For this adiabatic case, the parameter space for stable dust traps is reduced, compared to the isothermal case. We roughly explain this result as follows.  The boundary for pressure trapping (solid curve) expands to slightly larger widths. This expansion occurs because the pressure gradient from adiabatic bumps have an extra contribution from the temperature bump, in addition to the surface density contribution.  The more significant effect is that the RWI-stable region contracts, also moving to larger widths. 

This contraction of the RWI-stable region occurs for two reasons.  First, as just noted, with the additional temperature component the bump produces stronger pressure gradients and thus more shear. Second, the pressure gradient acceleration, $\Sigma^{-1} dP_{\rm H}/dR\propto \Sigma^{\Gamma-1}d\ln\Sigma/dR$, does not saturate with increasing $A_\Sigma$, but continues to increase since $\Gamma -1 = 1/3 > 0$.  Thus the amplitude-width curve of marginal stability does not steepen like the isothermal case discussed in Section \ref{subsec:ref}, or as seen in the left panel of Figure \ref{fig:eff_bumptypes}.  The net result of both shifts is a smaller region of parameter space for stable dust traps, when the bump is adiabatically heated vs.\ remaining isothermal.

The right panel of Figure \ref{fig:eff_bumptypes} considers a temperature bump with no surface density component (i.e. $A_T>0,\,A_\Sigma = 0$). The background disk parameters $n,q$ and $h_0$ are the same as those of the fiducial model.  In this case, there are no dust traps that are stable to the RWI. The pressure trapping boundary contracts significantly to narrower widths, compared to a surface density bump.  The effect arises because, with the scaling $P \propto \Sigma \sqrt{T}$, a \rev{temperature} bump produces weaker pressure gradients compared to a \rev{surface density} bump.
The RWI-stable region contracts, moving to wider widths.  The main effect is again that at large amplitudes the pressure gradient acceleration, $\Sigma^{-1} dP_{\rm H}/dR\propto dT/dR$, increases in amplitude (now $A_T$, instead of $A_\Sigma$) faster than either the isothermal or the adiabatic case, without any saturation. As a result, the marginal stability curve flattens to $A_T \propto W$ at large amplitudes.  The net effect of the shifts to both boundaries is that all bumps with a pressure maximum are in the RWI-unstable region.

The thermodynamics of any process that creates pressure bumps is crucial for understanding whether the dust traps can remain RWI-stable.  Isothermal pressure bumps have the largest parameter space of stable dust traps, which is reduced for adiabatically heated bumps.  A \rev{temperature} bump with no accumulation of surface density is unlikely to create a stable dust trap.  \rev{From Figure \ref{fig:eff_params} (right panel), lower $h_0$ values will introduce a region of stable dust traps for pure temperature bumps.  Nevertheless, by significantly reducing the allowed parameter space, our results disfavor the hypothesis of dust-trapping in gas temperature bumps.}
 
\begin{figure*}
    \centering
    \includegraphics[width=\textwidth]{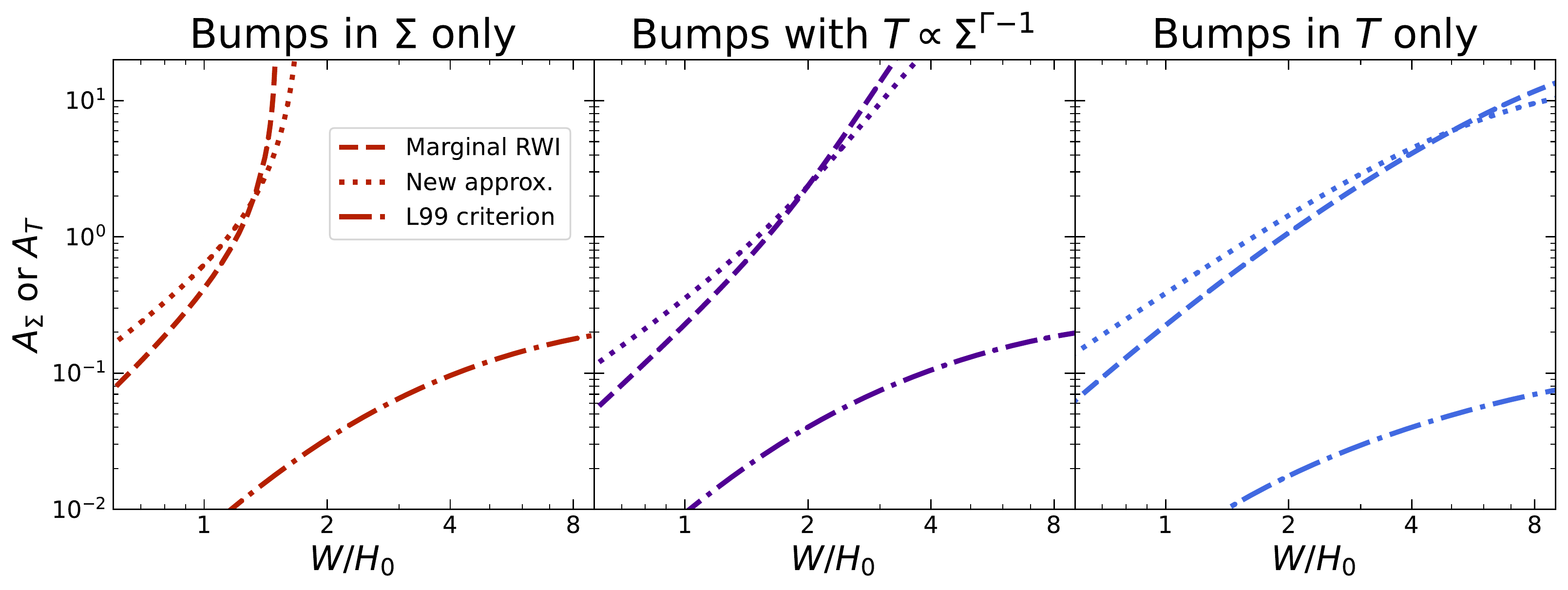}
    \caption{Our approximate stability criterion (\emph{dotted}) is compared to our numerical results (\emph{dashed}) for the same models as Figure \ref{fig:eff_bumptypes}. The approximate condition for marginal stability is $\min((\kappa^2+N_{R}^2)/\Omega_\mathrm{K}^2)=0.6$. The Lovelace criterion (\emph{dot-dashed}), a necessary but not sufficient criterion for the RWI, is shown for comparison.  Our approximate criterion can be used to estimate conditions for the RWI to occur.}
    \label{fig:new_criterion}
\end{figure*}

\subsection{New approximate stability criterion for RWI}\label{subsec:RWIstabapprox}


Unfortunately, there is no analytic criterion for the RWI which is both necessary and sufficient.  Such a criterion would greatly facilitate our exploration of stable dust traps, and many other applications of the RWI.  We report an approximate empirical criterion here, which might prove useful or spur further developments.  We first introduce some well-known stability criterion for context.

The Lovelace criterion is that a maximum in fluid vortensity gives a necessary, but not sufficient, criterion for the RWI \citepalias{lovelace_rossby_1999}. Figure  \ref{fig:new_criterion} confirms that the Lovelace criterion (dot-dashed curve) lies well below the numerically determined RWI stability boundary (dashed curve).  Recall that unstable (or potentially unstable in the case of the Lovelace criterion) regions lie above and to the left of stability boundaries.  

The Rayleigh criterion, $\kappa^2<0$, gives the axisymmetric condition for instability to radial oscillations for a barotropic rotating fluid, such as a disk \citep{chandra}. The generalization to baroclinic fluids is one of the   Solberg-H\o iland criterion, $\kappa^2+N_R^2<0$ \citep{tassoul}.  It is well known that the non-axisymmetric RWI occurs when disks are stable to both of these axisymmetric criteria \citepalias{lovelace_rossby_1999}.  In summary, the RWI criterion lies between the Lovelace and Solberg-H\o iland criterion.

We find a simple modification of the Solberg-H\o iland criterion, that somewhere in the flow:
\begin{gatheralign}
\kappa^2 + N_R^2 \lesssim 0.6 \Omega_\mathrm{K}^2\, .
\end{gatheralign}
This approximate condition gives an imperfect, but surprisingly good description of the numerically determined RWI criterion.  Physically, this criterion states that the squared radial oscillation frequency should be less than about 60\% of the squared Keplerian frequency, somewhere, for the RWI.

Figure \ref{fig:new_criterion} compares the new approximate (dotted curve) and precise numerical (dashed curve) criteria for the same isothermal, adiabatic and heated bump cases as Figure \ref{fig:eff_bumptypes}. The approximate criterion underestimates instability at low amplitudes and narrow widths, and overestimates instability in the opposite regime.  But as least on a logarithmic scale, accuracy is reasonable.

Thus our approximate explanations of the shape of the RWI stability boundary could be made more precise by a consideration of radial oscillation frequency, which is dominated by $\kappa$, with $N_R$ a modest correction in our models, and zero in the adiabatic case. With $\kappa^2= R^{-3}d (R^4\Omega^2)/dR$ (and $\Omega$ the orbital frequency including deviations from Keplerian due to pressure gradients), we justify that our arguments based on shear in $\Omega$ apply to the RWI.  We hope that our approximate criterion proves useful for similar interpretations or quick estimates, and especially that it might motivate deeper insights to the nature of the RWI.

\section{Conclusions}
\label{sec:discussion}
The leading hypothesis to explain the continuum rings imaged by ALMA is the trapping of dust in a disk bump with a pressure maximum.
This letter constrains this hypothesis by investigating whether these bumps can be stable to the RWI.
Regardless of their origin, the pressure bumps should remain dynamically stable for long enough to trap significant dust.  
We have shown that dust-trapping pressure bumps can be stable to the RWI, adding further theoretical support for the hypothesis.  Moreover,  our results could be used to plan and interpret searches for pressure bumps with ALMA, via the intensity and velocity shifts of molecular gas in the bumps.

Our stability analysis finds that low-amplitude pressure bumps cannot be stable dust traps. At low bump amplitudes, \rev{$A_\Sigma\lesssim0.2$ for our fiducial case}, the narrow widths needed for a pressure maximum also trigger the RWI.  For high enough bump amplitudes, however, stable dust traps exist for a range of bump widths that depends significantly on temperature. The temperature of the disk background and of the bump relative to this background are both important, especially with the background temperature parameterized as the disk aspect ratio, $h_0$.  

Cooler temperatures, in either bump or background disk, favor the existence of stable dust traps.  For lower values of the disk aspect ratio, $h_0$, stable dust traps are found for lower amplitude bumps and over a wider range of widths.  Our stability constraints thus imply that dust traps should be more readily produced in thinner, colder disks. 

Our analysis also constrains the allowed temperature of bumps relative to the disk background.  Cooler pressure bumps, i.e.\ those that are isothermal with the disk's background temperature, can be stable dust traps over a large range of bump widths, from one to several disk scale-heights.  As bump temperature increases, the range of stable widths decreases.  For hot pressure bumps, i.e.\ with no surface density excess, all bumps with a pressure maxima are unstable, for our fiducial disk model.

The background slopes of disk surface density and temperature are found to have a relatively modest effect on our results, over the relevant parameter range considered.  This finding limits the impact of uncertainties in disk parameters.  We also report a new approximate criterion for the RWI, that $\sqrt{\kappa^2+N_R^2}\lesssim0.75\Omega_\mathrm{K}$ somewhere in the flow.  

There are possible extensions that could also address some limitations of this initial study.  Our analysis of dust traps in gas bumps could be extended to dust traps at gap edges. 
Moreover the stability properties of gaps carved by planets could be analyzed (as in \citealp{lin2010, cimerman2023}).  Additional physical effects could be included such as 3D motions, radiative cooling, self-gravity of massive disks and magnetic fields.  See \cite{lesur2022} for a review of disk instabilities from these effects.  For the disk bumps considered here, the RWI (and the related \citealp{pp1985} instability for the barotropic case) is the instability that arises from the simplest and most general physical ingredients, and thus the natural starting point.  \rev{The effect of radiative cooling appears to be limited on the linear RWI (slightly decreases growth rates, \citealp{huang_rossby_2022}), even though it can affect the Rossby vortex lifetimes significantly \citep{fung_cooling-induced_2021}.
Nevertheless, more study is needed.}  Ultimately radiative transfer models, based on hydrodynamic numerical simulations with a distribution of dust grain sizes (as in \citealp{krapp2022}) could give more detailed and realistic observational predictions for ALMA.   

A key reason to better understand observed disk structures is to learn about how planets form.  Some dust traps may be caused by already-formed planets, and others may arise from (magneto)hydrodynamic processes in disks.  It is very important to understand which case is more prevalent.  However, in either case, dust that concentrates in these bumps is likely to grow into planetesimals.   Such growth could occur by enhanced collisional growth, direct gravitational collapse and/or dust concentration by the streaming instability \citep{chiang2010, johansen14}.  The streaming instability is a mechanism to create dust overdensities from the mutual aerodynamic coupling of dust and gas in disks \citep{yg05}.  These overdensities can then collapse gravitationally into planetesimals, typically in binary pairs \citep{nesvorny19}.  However, particle concentration by the streaming instability already requires locally elevated values of the dust/gas ratio \citep{jym09, li2021}.  This requirement can become even more stringent when a broad dust size distribution is accounted for \citep{krapp2019}.  Several studies show that the streaming instability is most likely to be triggered in overdense dust rings, caused by ice lines and/or pressure bumps \citep[e.g.\ ][]{draz2013, Schoon2017, draz2018,ida2021}.  And the streaming instability has been studied in the specific context of pressure bumps \citep{onishi2017, carrera_protoplanetary_2021}.  A better understanding of the dust structures observed by ALMA is thus of crucial importance for theoretical models of planet formation.

\medskip
\noindent
The authors acknowledge support from NASA through TCAN grant 80NSSC21K0497, and thank TCAN team members, including Wladimir Lyra, Chao-Chin Yang, and Jacob Simon for constructive comments.
L. K.  acknowledges support  by the Heising-Simons 51 Pegasi b postdoctoral fellowship.  The authors are thankful to Feng Long, Ilaria Pascucci and the Star and Planet Formation Theory Group at Steward Observatory for useful discussions and feedback.


\bibliographystyle{aasjournal}
\bibliography{bibliography}



\end{document}